\documentclass[prl,aps,superscriptaddress,showpacs,twocolumn,floatfix]{revtex4}
\usepackage{amsmath}
\usepackage{amssymb}
\usepackage{xspace}
\usepackage{graphicx}

\newcommand{\physrev}{Phys. Rev.}

\newcommand{\epjb}{Eur. Phys. J. B}

\newcommand{\ssc}{Solid State Commun.}
\newcommand{\zetf}{Zh. Eksp. Teor. Fiz.}
\newcommand{\jetp}{Sov. Phys. JETP}
\newcommand{\jap}{J. Appl. Prob.}

\newcommand{\ignore}[1]{}
\newcommand{\etal}{\emph{et al.}\xspace}
\newcommand{\ie}{\emph{i.e.}\xspace}
\newcommand{\beq}{\begin{equation}}
\newcommand{\eeq}{\end{equation}}
\newcommand{\bea}{\begin{eqnarray}}
\newcommand{\eea}{\end{eqnarray}}
\newcommand{\lb}{\overline{l}}
\newcommand{\kB}{k_\mathrm{B}}
\newcommand{\Rmax}{R_\mathrm{max}}

\begin{document}

\title{Distribution of the resistance of nanowires with strong 
impurities}

\author{Christophe Deroulers}
\email{cde@thp.uni-koeln.de}
\affiliation{Institut f\"ur Theoretische Physik, Universit\"at zu 
K\"oln, Z\"ulpicher Str. 77, 50937 K\"oln, Germany}

\date{\today}

\begin{abstract}

Motivated by recent experiments on nanowires and carbon nanotubes, we
study theoretically the effect of strong, point-like impurities on the
linear electrical resistance $R$ of finite length quantum wires. Charge
transport is limited by Coulomb blockade and cotunneling. $\ln R$ is
slowly self-averaging and non Gaussian. Its distribution is Gumbel with
finite-size corrections which we compute. At low temperature, the
distribution is similar to the variable range hopping (VRH) behaviour
found long ago in doped semiconductors. We show that a result by Raikh
and Ruzin does not apply. The finite-size corrections decay with the
length $L$ like $1/\ln L$. At higher temperatures, this regime is
replaced by new laws and the shape of the finite-size corrections
changes strongly: if the electrons interact weakly, the corrections
vanish already for wires with a few tens impurities.

\end{abstract}

% 73.63.-b Electronic transport in nanoscale materials and structures
%         (see also 73.23.-b Electronic transport in mesoscopic systems)
% 73.23.Hk Coulomb blockade; single-electron tunneling
% 72.20.Ee Mobility edges; hopping transport

\maketitle

 Understanding the charge transport in materials with impurities has
been a long standing problem which attracts new interest because of
recent experiments on systems of nanoscopic size~\cite{gao-et-al}. In
the past, the case of doped semiconductors has been extensively
studied~\cite{livre-shklovskii-efros}. Because of impurities, the
electrons are, in equilibrium, in localized states. These states are
randomly spread in energy and space. At low temperature, conduction
results of a sequence of thermally activated hops of the charge
carriers, which tunnel from one state to another, borrowing the energy
difference from the phonons bath if needed. The spatial range of these
hops results from a balance between tunneling (favoring short ranges)
and activation (finding a state close to the Fermi level is more likely
if the range is long). As a consequence, the low-field resistance $R$ of
amorphous semiconductors follow Mott's VRH law, $\ln R \sim
T^{-1/(D+1)}$ in dimension $D$, with peculiarities in the special case
of 1D~\cite{kurkijarvi, lee}. In presence of long-range Coulomb
repulsion, as for crystalline semiconductors, the states close to the
Fermi surface are rare and this law must be replaced at very low
temperatures by the Shklovskii-Efros VRH law, $\ln R \sim
T^{-1/2}$~\cite{livre-shklovskii-efros}.

 Here we deal with 1D systems of nanoscopic cross-sections (nanowires)
with point-like, strong impurities, and repulsive interactions between
the electrons, at low applied electric field. In a pure nanowire, or
between two impurities, the physics is described by the
Tomonaga-Luttinger liquids (LL) model rather than a Fermi liquid
because, in a 1D fermionic system, the excitations have a bosonic
character. The case of a few strong impurities, or of a uniform density
of weak impurities, have been
studied~\cite{kane-fisher-furusaki-nagaosa, livre-giamarchi}. $R$ was
found to behave like powers of $T$ at least in some regimes, with
exponents which depend on the interaction parameter $K$ of the LL. As
was recently shown~\cite{fogler-m-n}, in the case of many strong
impurities, Coulomb blockade~\cite{gao-et-al} and statistical effects
yield different laws: at low temperature, a VRH-like regime is
predicted, while at medium temperature a new, more complicated law
(which can, on a small range of $T$, look like a power law) should
replace it.

 In this Letter, we first confirm numerically the existence of these two
regimes for the average of $\ln R$. But they might be hard to
distinguish due to huge sample to sample fluctuations. Therefore, we
study the full distribution of $\ln R$. $R$ and $\ln R$ are very slowly
self-averaging and non Gaussian. Unless the wires are very long, the
distribution of $\ln R$ is Gumbel with corrections that decay like
$1/\ln L$ at low $T$. They are much weaker at medium $T$ and have a
different shape. We think this will help uncover experimentally the two
regimes. Finally, we discuss the ergodicity hypothesis often used in
experimental studies.

 Our plan is the following: after introducing the model and the
simulation technique, we show numerically that each sample has a high
$T$ regime where its resistance is proportional to the length and a
``non extensive'' regime where the voltage drop occurs essentially on a
few \emph{breaks}. We check the validity of a percolation-like
approximation to $R$. Then we give our conclusions for the average of
$\ln R$ and derive the distribution of $\ln R$.

\emph{Model.} Our model is similar to the one of
Refs.~\cite{nattermann-g-ld, fogler-t-s-malinin-n-r}. We consider
spinless electrons. Their interactions are short range because of
screening by a nearby gate, which gives a finite capacitance per unit
length $C$ to the nanowires. Each wire is cut into a series of weakly
coupled quantum dots by $N$ impurities. They are placed randomly
according to a Poisson process with mean spacing $\lb$ large compared to
the Fermi wavelength $\lambda_F$. In the quasiclassical limit, which is
relevant because the impurities are strong, each dot $j$ can only have
an integer number, $q_j$, of electrons. Thus the charge transport
proceeds by discrete hops, like in disordered semiconductors, and is
limited by activation and tunneling. But unlike the localized states in
the VRH model, here the energies of the states are correlated with their
positions. Indeed, in the ground state, $q_j$ is the closest integer
$q_j^{(0)}$ to the mean charge $Q_j=l_j/\lambda_F + \mu/\Delta_j$, where
$\mu$ is the chemical potential and $\Delta_j = 1/(C l_j)$ is the
charging energy (we use $e=\kB=1$). The energy of dot $j$ is
$E_j(q_j)=(q_j-Q_j)^2 \Delta_j/2$~\cite{livre-giamarchi}. The typical
charging energy $\Delta := 1/(\lb C)$ is the natural energy scale of the
problem. In addition, each charged excitation $q_j \ne q_j^{(0)}$ may
itself be neutrally excitated by an energy difference $\sim K \Delta_j$.

 Having in mind experiments where a four-point measurement is
performed~\cite{gao-et-al}, we do not pay attention to the contacts
between the wire and the two leads. We assume for convenience that each
impurity has the bare transparency $1/(e\pi) \approx 0.12$. According
to~\cite{larkin-lee, fogler-t-s-malinin-n-r, artemenko-remizov}, because
of the LL effects, its \emph{effective} tunneling transparency may be
found by an instanton calculation, yielding $\exp(-s_{j,j+1})$ with
$s_{j,j+1} = 1 + (K^{-1}-1) (\sigma_{j-1} + \sigma_j)$ with $\sigma_j =
\ln [K/(C \lambda_F)/\max(K \Delta_j, T) ]$. The tunneling process is
limited by neutral excitations at low $T$ and by thermal fluctuations at
high $T$. This extends readily to the probability $\exp(-s_{j,k})$ of
the cotunneling of $k-j$ electrons, each through one of the impurities
$j, \ldots, k-1$ (which is formally equivalent to the hop of one
electron from the dot $j$ to the dot $k$). At low temperatures,
cotunneling allows to shortcut the narrow dots with high charging energy
--- this is why $\ln R$ increases slower than $1/T$ as $T$ goes to 0.

 Putting things together, we find the probability that an electron hops 
from dot $j$ at $x_j$ with initial charge $q_j$ to dot $k$ at $x_k$ with 
initial charge $q_k$ under field $F$: \beq \label{expression_w} 
w_{j,q_j,k,q_k} = e^{-s_{j,k}} \gamma_0 E \left[ \exp(E/\kB T) - 1 
\right]^{-1} \eeq where $E = E_j(q_j-1) - E_j(q_j) + E_k(q_k+1) - 
E_k(q_k) - e F (x_k-x_j)$. The energy difference $E \gtrless 0$ is 
provided/absorbed by the phonons bath; the field $F$ favors one hopping 
direction. $\gamma_0$ depends on the coupling with 
phonons~\cite{miller-abrahams}.

 In a steady regime, the average occupation probability of each state 
should be constant, therefore we write down for each state a balance 
equation between incoming and outgoing charge flows. Solving this in a 
general out-of-equilibrium setting is difficult, but, following the idea 
of Miller and Abrahams~\cite{miller-abrahams}, we restrict to the small 
$F$ regime and simplify these equations to the first order in $F$. The 
resulting equations are the Kirchhoff's laws for a resistor network 
between nodes labeled by $j,q_j$. In the following we are concerned with 
the low temperature regime $T \lesssim 10 \Delta$ and we assume that 
only the first two charged excitations on each dot play a role. If this 
may not be true for all dots, it will certainly be the case for the dots 
which contribute the most to the resistance by their high charging 
energy. Moreover, we further simplify~\cite{fogler-m-n} to a network 
with one node per dot $j$ and resistances $R_{j,k}$ equal to the minimum 
over $q_j$ and $q_k$ of the resistances between the states $j,q_j$ and 
$k,q_k$. In this network, each node is connected to all others, although 
connections between remote dots are likely to be very resistive.

\emph{Simulations.} We performed simulations by drawing at random the 
positions of the impurities of 10000 wires for each value of $K$, 
impurities number $N$ and mean spacing $\lb$. The low field resistance 
$R$ of each wire is computed as the resistance of the equivalent network 
defined above; more precisely, we took care to manipulate $\ln R$ rather 
than $R$ to prevent overflows at low $T$ where $R$ become huge; 
arbitrary precision arithmetics was used when needed to extract 
subdominant values. Simple linear solving of the Kirchhoff's equations 
fails at low temperatures because of numerical instabilities, but 
repeatedly eliminating one node of the network by applying a generalized 
star-triangle transformation~\cite{dyre} until only two nodes are left 
is efficient and stable.

 For each wire, we also computed the resistance of the ``best path'',
$R_{bp}$, which is the less resistive of the subnetworks without loops
connecting the nodes 0 and $N+1$ and going always in the same direction,
in analogy with the percolation approach to
VRH~\cite{livre-shklovskii-efros}. A link in the best path is termed a
\emph{hop}. We found that $\ln R_{bp}$ is a very good approximation of
$\ln R$ up to $T \sim 10 \Delta$. Clearly, $R_{bp} > R$ because the
full network has derivations around the best path. Empirically, $\ln
R_{bp} / \ln R < 1.01$ always, and this ratio goes to 1 as $T$
decreases. On our curves the averages of $\ln R_{bp}$ and $\ln R$ can't
be distinguished. In the following we use $R_{bp}$ for our analytical
discussion because it is a \emph{sum} of individual resistances.

\begin{figure} 
\includegraphics[width=0.453\textwidth]{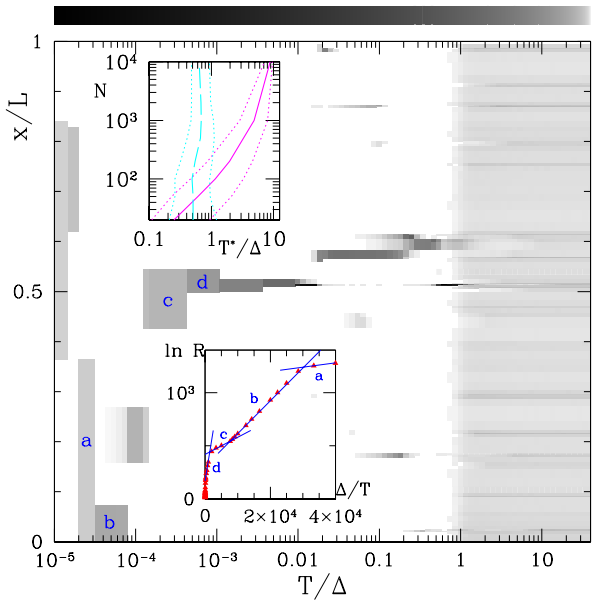}

\caption{\label{figure-gradient} The gradient of the electrical
potential $V(x)$ along the best path in one random sample with $N=100$
impurities, interaction parameter $K=0.3$ and mean impurities spacing
$\lb=10^3\lambda_F$. For each value of the temperature $T$, we apply a
voltage unity to the wire and we compute the gradient $V'$ of $V$ on
each hop along the best path (see text). Then we color this hop with a
grey intensity depending on $V'$ (the bar above the figure indicates the
intensity coding: white is null gradient and black is the maximal
gradient over the whole figure). \textbf{Top inset:} The sum of the
squares of the voltage drop of each hop along the best path is 1 at low
temperature (one hops dominates) and $1/N$ at high temperature ($N$ hops
with drop $1/N$). For each of $10^4$ samples from the same distribution,
we look for the temperature $T^*$ where it is equal to $1/\sqrt{N}$. 
This separates the non-extensive and extensive $R$ regimes (see text). 
Strong lines: median value of $T^*$ for several $N$ values (solid:
$K=0.1$, dashed: $K=0.7$). Dotted lines: boundaries of the 68\%
confidence interval on $T^*$ at a given $N$. \textbf{Bottom inset:} Plot
of $\ln R$ w.r.t. $1/T$. The curve is made of activated segments. In
each one, the voltage drop occurs in one break which is identified by a
letter on the main plot.} \end{figure}

 From the simulations we conclude first that each sample has two well
separated regimes: at high $T$ the voltage drop is almost uniform and
the resistance $R_{bp} \approx R$ is extensive (proportional to the
length), but at low $T$ the voltage drops mainly in one or a few
regions, the \emph{breaks} (Fig.~\ref{figure-gradient}), which makes $R$
non extensive unless the wire is extremely long~\cite{raikh-ruzin}. The
positions of the breaks strongly depend on $T$ and their width increases
as $T$ goes to zero and cotunneling through more and more impurities
takes place. Fig.~\ref{figure-gradient} shows this for one particular
sample but it is true for all but pathological samples and for all
values of $K$. The very low temperature regime $T \le 0.01 \Delta$ is
cut into sample-specific ranges where $R(T)$ is activated, as was
observed experimentally~\cite{gao-et-al}: in each range of temperature,
a different break fixes the voltage drop.

\begin{figure}
\includegraphics[width=0.36\textwidth]{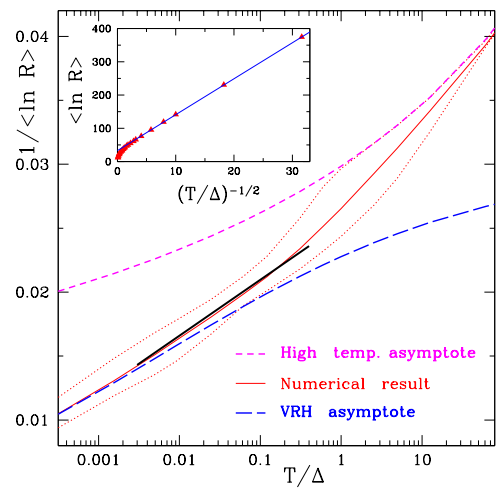}

\caption{\label{figure-moyenne-lnR} Solid line: numerical results for 
the inverse of the average of $\ln R$ over $10^4$ wires similar to the 
one of Fig.~\ref{figure-gradient} as a function of $T/\Delta$. The error 
bars are smaller than the line's thickness. The thin, dotted lines above 
and below indicate the boundaries of the 68\% confidence interval at 
each temperature. Strong short-dashed line: high $T$ asymptote and upper 
bond, reached when the electrons tunnel sequentially through all 
impurities. Strong long-dashed line: low $T$ asymptote where $\ln(R)$ 
follows the VRH law, as found from the inset. The thick straight line 
indicates the validity of the new law predicted in~\cite{fogler-m-n} in 
some range of $T$. \textbf{Inset.} Dots: plot of the average $\ln R$ 
w.r.t. $\sqrt{\Delta/T}$ to show that, at low $T$, the VRH law is 
obeyed. The error bars are smaller than the symbols' size. Straight 
line: linear fit , which is used on the main plot as VRH asymptote.} 
\end{figure}

 The second conclusion is a confirmation of the results
of~\cite{fogler-m-n} for the average over many samples $\langle \ln R
\rangle$: a VRH-like law $\langle \ln R \rangle \sim 1/\sqrt{T}$ holds
at low $T$ but has to be replaced by a new regime at intermediate $T$
where $1/\langle \ln R \rangle$ is a affine in $\ln T$
(Fig.~\ref{figure-moyenne-lnR}). However, unless the wire is very long,
the temperature range where this new, non VRH-like law can be observed
is small and it is in the cross-over between low and high temperatures
(cotunneling and sequential tunneling), so that distinguishing the two
laws is hard, all the more than the sample to sample fluctuations are
large (Fig.~\ref{figure-moyenne-lnR}). But we shall see that looking at
the full distribution of $\ln R$ may help.

\begin{figure}
\includegraphics[width=0.36\textwidth]{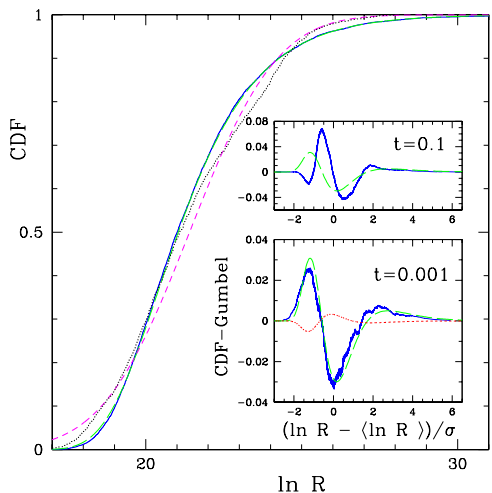}

\caption{\label{figure-distribution-lnR} Solid line: CDF of $\ln R$ for
wires with $N=100$ impurities, $K=0.7$ and $\lb/\lambda_F=1000$ at
temperature $T=0.1 \Delta$ from our simulation. Dotted line: the same
for one sample but $\mu$ varying uniformly in the rage $-1$..$1$ to show
that the ergodic hypothesis is only approximate. Long/short dashed line:
Gumbel/Gauss distributions having the same average and variance than
this CDF. The Gumbel distribution is almost superimposed on the data. 
\textbf{Insets.} Solid line: finite-size effects for the same wires but
$K=0.3$ and at $T=0.1 \Delta$ resp. $0.001 \Delta$ (difference between
the CDF of the centered and reduced $\ln R$, and a Gumbel distribution).
Long dashes: eq.~(\ref{fsc-vrh}) with $n=100$. Dots: shape of the
finite-size effects after Raikh and Ruzin's result.} \end{figure}

\emph{Analytical results for the distribution of $\ln R$.} From now on,
we restrict to the low $T$ regime, where breaks are observed. Taking for
simplicity uniform $s$ and $\lambda_F=0$, and neglecting correlations
between hops, the probability that a given hop on the best path has $\ln
R < u$ is \beq P\{ R < e^u \} = 1 - \prod_{j=1}^k f[ \frac{T}{\Delta}
(\ln u-j \, s)] \ \textrm{with}\ k = \lfloor u/s \rfloor \eeq
where~\cite{fogler-m-n} $f(x) := 1 - x \left[ 1 - \exp(-1/x) \right]$.
The number $n$ of hops in a sample is proportional to $N$ (at low $T$,
$n \approx \sqrt{s T/\Delta} N$) and, because of the equivalence of
statistical ensembles, we can replace $n$ or $N$ by $L$ in our
asymptotic estimations. The most resistive hop has typically a
resistance $\Rmax$ such that $P\{ R < \Rmax \}^n \approx 1/2$. As long
as $\Rmax$ is large compared to the average hop resistance times
$n$~\cite{benarous}, we are in the non extensive $R$ regime, $R_{bp}
\approx \Rmax$ --- this regime extends at least to wires with several
tens of thousands of impurities. If $R_{bp}/s \gtrsim 6$, $- \ln P [R >
e^u]$ is well approximated by an integral, which itself takes two
forms~\cite{fogler-m-n}: (a) $T u^2/(s \Delta)$ if $u \, T/\Delta \ll 1$
and (b) $(u/s) \ln(u \, T/\Delta)$ if $u \, T/\Delta \gg 1$. The
conditions $R_{bp}/s \gg 1$ and $T/\Delta \ln R_{bp} \ll 1$ are always
fulfilled at (very) low temperatures, say $T \le 0.001 \Delta$. 
Furthermore, we verified numerically that the cumulative distribution
function (CDF) of $\ln R$ of the hops along the best path of the wires
(with nonuniform $s$) is in very good agreement with the shape
$\exp(-u^2)$ once rescaled. Then $\ln R \approx \ln R_{bp}$ has for
large $n$ a Gumbel distribution because it is the maximum of $n$ random
variables with exponential decay of the probability
density~\cite{livre-david, huillet}. Its average behaves like $\sqrt{s
(\ln n) \Delta/T}$ and its standard deviation $\sigma$ like
$a/2/\sqrt{(\ln n)T/(\Delta s)}$ where $a:= \pi/\sqrt{6}$, which is
similar to the results of Lee \etal for the VRH model~\cite{lee} and
agrees with the low $T$ result of~\cite{fogler-m-n}. Both $R$ and $\ln
R$ are non gaussian but self-averaging. We prefer to study $\ln R$
because it is easier to sample numerically. We computed the CDF of the
centered and reduced $\ln R$ with finite $n$
corrections~\cite{hall-racz}: \bea \label{fsc-vrh} & & P\{\frac{\ln R -
\langle \ln R \rangle}{\sigma} < x\} = \exp[-g(x)] \{ 1 - \\ & &
\nonumber g(x) [ \pi^2/24 (1-x^2) + x \zeta(3)/(2 p) ] / \ln n +
\mathcal{O} [ 1/(\ln n)^2 ] \} \eea where $g(x)=\exp(-\gamma-a\,x)$,
$\zeta(3) \approx 1.202$ and $\gamma \approx 0.577$. This expression is
successfully compared to the numerical CDF of $\ln R$ in the inset of
Fig.~\ref{figure-distribution-lnR}. Since $1/\ln n$ goes slowly to zero,
the actual $\ln R$ of each sample will likely lie away from the average
$\langle \ln R \rangle$ even if $n$ gets large. This is why activated
segments are still visible on long wires.

 Pursuing the work of Lee \etal, Raikh and Ruzin~\cite{raikh-ruzin}
evaluated the distribution of $\ln R$ in the case of 1D VRH. 
Interestingly, their result tends to a Gumbel law for very short wires
or at very low $T$. But the finite-size effects are in $1/N^2$ and do
not match eq.~(\ref{fsc-vrh}), nor our numerical results (bottom inset
of Fig.~\ref{figure-distribution-lnR}).

 At not so low temperatures, say $T > 0.01 \Delta$, the CDF of
individual hops doesn't agree with form (a) above but, if interactions
are not too strong, with form (b). The VRH-like law $\langle \ln R
\rangle \sim 1/\sqrt{T}$ is replaced by $\langle \ln R \rangle \approx s
\ln n/\ln[(s T/\Delta) \ln n]$ where the number of hops $n$ is
proportional to $L$ as before. The difference between the averages in
the two regimes is small compared to the fluctuations $\sigma \sim 1/\ln
\ln L$. But a strong change in the finite-size effects of the CDF of
$\ln R$, which is Gumbel-like, should be easy to detect: their amplitude
is reduced by $\ln [(s T/\Delta) \ln n]$ \ie more than 3 in our data and
their shape is changed. Fig.~\ref{figure-distribution-lnR} shows that,
for $K=0.7$ and only $N=100$ impurities one can't tell the difference
between the CDF of $\ln R$ and a Gumbel CDF. For stronger interactions
(smaller $K$), say $K=0.3$, a typical break has only a few impurities,
(b) is not valid, and the CDF is hard to characterize, but the
deviations from a Gumbel law can't be confused with the
formula~(\ref{fsc-vrh}) of the VRH case (top inset of
Fig.~\ref{figure-distribution-lnR}) and might still be used as an
evidence of strong impurities in a LL wire.

 In an experimental study of the distribution of $\ln R$, it is common
to measure $R$ for the same sample but several values of the gate
potential: this changes $\mu$, and the activation energies strongly
depend on $\mu$. This was argued to yield the same results as using many
samples (see \cite{hughes-et-al} and references therein). However, in
addition to varying $\mu$, we recommend to use as many samples as
possible because we found numerically that this ergodicity hypothesis is
only approximate and may lead to systematic errors
(Fig.~\ref{figure-distribution-lnR}).

{\bf Acknowledgments.} 

We acknowledge financial support from the SFB 608. It is a pleasure to
thank M.~Fogler, T.~Nattermann and Z.~R\'acz for discussions and
support.

\end{document}